\documentclass[aps,prb,twocolumn]{revtex4}
\usepackage{amssymb}
\usepackage{graphicx}

\begin{document}
\author{Kavita Mehlawat and Yogesh Singh}
\affiliation{Indian Institute of Science Education and Research Mohali, Sector 81, S. A. S. Nagar, Manauli PO 140306, India}

\date{\today}

\title{First order density wave like transitions in surface doped Na$_2$IrO$_3$}

\begin{abstract}
We demonstrate that the surface of the honeycomb lattice iridate Na$_2$IrO$_3$ is extremely tunable by plasma etching.  We have succeeded in turning the surface of Na$_2$IrO$_3$ metallic by Ar plasma etching which leads to the removal of Na from the surface.  The surface structure does not change in this process as revealed by grazing incidence small angle x-ray scattering (GISAXS).  The sheet resistance $R_s$ can be reduced by several orders of magnitude by varying the etching duration.  Temperature dependent $R_s(T)$ for the metallic samples show signatures of spin or charge density wave transitions with abrupt changes in $R_s$.  Thermal hysteresis between cooling and warming measurements across the transition indicates a first order transition.  For the most metallic sample $R_s(T)$ data at low temperatures follow a $T^2$ behaviour suggesting normal Fermi liquid behaviour.  
 
\end{abstract}

\maketitle
Doped Mott insulators can show exotic physics like high temperature superconductivity \cite{Lee2006, Tokura1990}, heavy Fermion behaviour  \cite{Carter1991,Tokura1993}, or other correlated electron behaviour \cite{Imada1998}.  Some aspects of the low energy physics of the high temperature Cuprate superconductors can be understood within a half filled single band Hubbard model \cite{Lee2006, Pickett1989}.  Recently many iridate materials have been found to show Mott insulating behavior where the low energy physics can also be described by a half filled single band model \cite{Kim2008, Jackeli2009, Kim2009}.  It has been suggested that doped iridate Mott insulators could be avenues to search for high temperature superconductivity \cite{Wang2011}.  Sr$_2$IrO$_4$ is a specially attractive material because it has the same crystal structure as La$_2$CuO$_4$ \cite{Crawford1994}, and many of its magnetic properties are similar to the cuprates \cite{Crawford1994, Kim2012}.  Electron doping of Sr$_2$IrO$_4$ has been successfully achieved by creating oxygen deficient single crystals Sr$_2$IrO$_{4-x}$ \cite{Korneta2010}.  While these samples become metallic and show significant changes in their magnetic properties, no superconductivity has thus far been observed \cite{Korneta2010}.  Recently in situ surface electron doping by potassium deposition has also been achieved and it was found that Fermi arcs, pseudogap, and a low temperature d-wave gap exist for these samples demonstrating properties in complete analogy with the cuprates \cite{Kim2014, Kim2016, Yan2015}.  The more traditional signatures of superconductivity like zero resistance state and the Meissner effect have still not been observed for doped Sr$_2$IrO$_4$.  Nevertheless, the strong dependence of electronic and magnetic properties of Sr$_2$IrO$_4$ with small doping suggests that properties of iridates could be highly tunable. 

Recently, the family of layered honeycomb lattice iridates $A_{2}$IrO$ _{3} (A =$~Na, Li) has garnered a lot of attention.  In these materials spin-orbit entangled effective moments $J_{eff} = 1/2$ sit on a honeycomb lattice leading to novel magnetic properties\cite{Jackeli2009, Chaloupka2010, Singh2010, Choi2012, Singh2012}.  Na$_{2}$IrO$ _{3}$ which is electrically insulating with a band gap of 350~meV ~\cite{Comin2012} shows frustrated magnetism as evidenced by a long range magnetic ordering temperature $T_N = 15$~K which is much reduced compared to the Weiss temperature $\theta = - 120$~K \cite{Singh2010}.  There is now direct evidence of dominant bond directional exchange interactions in Na$_{2}$IrO$ _{3}$ suggesting that Kitaev like exchanges maybe dominant in this system \cite{Chun2015}.   
 
Unconventional spin-triplet superconductivity and topological superconductivity have been predicted to emerge with doping in the Kitaev-Heisenberg model on the honeycomb lattice \cite{You2012, Hyart2012, Okamoto2013, Scherer2014, Liu2015}.   Additionally, spin and charge density wave, spin/charge bond-order, and electronic dimerization instabilities have been predicted with varying doping \cite{Scherer2014}.  The honeycomb lattice ruthenates and iridates are already close to coupled structural, magnetic, and orbital instabilities as evidenced by Ru-Ru dimerization in Li$_2$RuO$_3$ \cite{Miura2007} and in $A_2$IrO$_3$ ($A =$Li, Na) under small pressures\cite{Clancy}.

In this work we report the discovery that the surface of Na$_2$IrO$_3$ crystals turns metallic when we expose it to high energy Ar Plasma.  With varying exposure times ($0$--$30$~minutes) we can enhance the conductivity by several orders of magnitude and go from an insulating surface to a metallic one.  The samples etched for $5$~minutes remain insulating although the charge gap is reduced.  The $10$~minute etched samples show behaviour between that of insulators and metals with a weak increase in resistance on cooling below $300$~K before becoming temperature independent below about $200$~K and having finite values as $T \rightarrow 0$.  The metallic Na$_2$IrO$_3$ samples obtained for larger than $10$~minutes plasma etching show exotic first order phase transitions reminiscent of spin/charge density wave (S/CDW) or structural transitions.  Specifically the $20$~minute sample shows an abrupt increase in the sheet resistance $R_s$ at $T_o = 220$~K while cooling.  This DW-like transition occurs at $230$~K while warming.  Surprisingly, the magnitude of the increase in $R_s$ at $T_o$ is magnetic field dependent suggesting that spin degrees of freedom are involved.  For the most metallic $30$~minute sample, around $T = 95$~K we observe a step-transition below which $R_s$ falls by more than an order of magnitude.  This is also a first order transition as we observe a hysteresis of $\approx 10$~K between the cooling and warming data.   For the $30$~minute etched samples the $R_s(T)$ data below $T =10$~K could be fit to a $T^2$ behavior which suggests normal Fermi liquid behavior.  These S/CDW-like transitions on doping Na$_2$IrO$_3$ are consistent with recent predictions \cite{Scherer2014}.

The single crystalline Na$_2$IrO$_3$ were synthesized as described elsewhere \cite{Singh2010}.  The surface of the plate-like crystals were modified by bombarding a freshly cleaved surface with high energy Argon (Ar) plasma for varying amounts of time ranging from $0$ to $40$~minutes.  The details of the parameters used for this plasma treatment is given in Table~\ref{Table-etch-parameters}.  The surface structure of the crystals before and after the plasma treatment was measured using grazing incidence small angle x-ray scattering (GISAXS).  The chemical composition of the sample surface before and after the plasma treatment was checked using energy dispersive x-ray (EDX) analysis with a JEOL scanning electron microscope (SEM). The electrical transport was measured using a Quantum Design physical property measurement system in the temperature range $2$~K to $305$~K\@.    

\begin{table}
\caption{Parameters from reactive-ion etching  }
\begin{ruledtabular}
\begin{tabular}{|c|c|}
Flow of Ar gas  & 80~SCCM   \\
Chamber Pressure & 80~mTorr   \\
RF Power  &  200~W   \\
RF Bias voltage &  -500~V  \\
Temperature & 10 - 20~$^\circ$C   \\
\end{tabular}
\end{ruledtabular}
\label{Table-etch-parameters}
\end{table}

Before we present our results we would like to note the following points.  The etched surfaces degrade in lab atmosphere.  The changes in the transport properties on etching are thus temporary and revert back to their original behaviour after exposure of the etched surface to lab atmosphere.  The time taken for the etched surfaces to degrade and revert back to insulating behaviour is about 1--2 days for the samples etched for $10$~minutes.  This time gets shorter for longer etching times.  The samples etched for $40$~minutes degrade in about an hour.  The transport of the crystal surface opposite to the etched surface does not change and remains highly insulating.  Additionally, if a fresh surface was exposed by cleaving off the etched surface, it showed insulating behaviour similar to undoped Na$_2$IrO$_3$.  These observations suggest that varying depths of the Na$_2$IrO$_3$ are affected with varying etching times and most likely only a small depth close to the top surface is modified.
  
\begin{figure}[h]   
\includegraphics[width= 3 in]{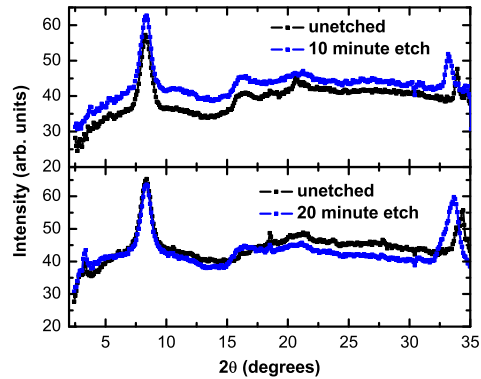}    
\caption{(Color online) A comparison of grazing incidence small angle x-ray scattering patterns for Na$_2$IrO$_3$ before and after varying periods of etching using an Ar plasma.
\label{Fig-gaxs}}
\end{figure} 

We have looked at the surface structure of Na$_2$IrO$_3$ using grazing incidence small angle x-ray scattering (GISAXS) measurements before and after the etching.  The depth $D$ probed in GISAXS measurements is given as $D = d~Sin\theta$, where $d$ is the attenuation length for this compound at the given x-ray energy, and $\theta$ is the incidence angle measured from the surface.  We estimate $D \approx 20$~nm with $d \approx 1.25 \times 10^{−3}$~cm at X-ray energy $15$~keV and $\theta = 0.1^o$.  Representative GISAXS patterns of two samples before and after the plasma etching are shown in Fig~\ref{Fig-gaxs}.  It is known through scanning tunneling microscopy (STM) measurements that the surface structure of Na$_2$IrO$_3$ is different from the bulk due to surface reconstruction \cite{Luepke2015}.  Nevertheless, a comparison of the GISAXS patterns before and after the plasma treatment suggests that the overall surface structure does not change after etching.  The most significant change is for the peak at high angles ($33$--$34^o$) which shifts to slightly smaller angles after etching suggesting that the cell parameter contributing to this peak increases.  However, there is no systematic evolution of the peak position with etching time.  For all etched samples, the peak position is approximately the same and smaller (in angle) than the corresponding peak in the unexposed Na$_2$IrO$_3$ surface by $\approx 0.5^o$.

Chemical analysis using energy dispersive spectroscopy on several spots of the crystals before and after the plasma etching gave the average chemical composition given in the Table~\ref{Table-EDS}.  From these results it is clear that Na is progressively being removed with increasing etching times.  Therefore, the plasma etching leads to hole doping (of the surface at least).       
 
\begin{table}[h]
\caption{Average ratio of desired elements from Energy dispersive x-ray analysis}
\begin{ruledtabular}
\begin{tabular}{|c|cc|}
Exposure time(min.) & Average Na  & Average
Ir \\ \hline
0 & 1.76 & 1   \\
10 &  &    \\
20 &  &  \\
30 &  1.62 & 1  \\
40 &  1.56 &  1  \\
\end{tabular}
\end{ruledtabular}
\label{Table-EDS}
\end{table}
   
Since the thickness of the surface layer affected by the etching is unknown and most likely depends on the etching time, we present electrical transport as sheet resistance $R_s$ given in the units $\Omega/$sq.  A sheet resistance of $1\Omega/$sq means that a square sheet will have a resistance of $1\Omega$ regardless of the size of the square.  

\begin{figure}[t]   
\includegraphics[width= 3 in]{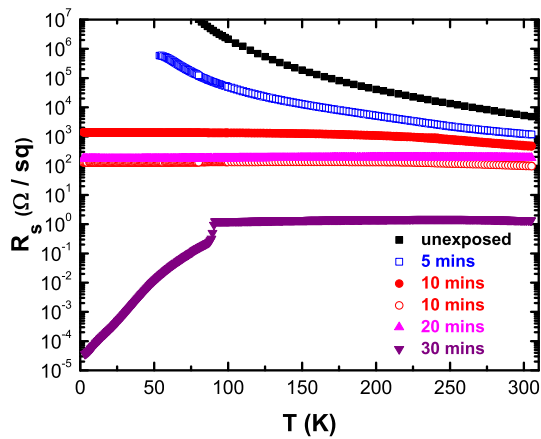}    
\caption{(Color online) A semi-log plot of the sheet resistance $R_s$ versus temperature $T$ for Na$_2$IrO$_3$ after varying periods of etching using an Ar plasma.
\label{Fig-all}}
\end{figure}

Transport for the sample etched for $40$~minutes could not be measured as the sample surface degraded within an hour before contacts could be made and cured.  The sheet resistance $R_s$ versus temperature $T$ for all other samples are shown on a semi-log scale in Fig.~\ref{Fig-all} to highlight the change in $R_s$ by several orders of magnitude with increasing etching times.  The sample exposed for $5$~minutes remains insulating although the band gap reduces to $\approx 1400$~K compared to $\approx 3600$~K for Na$_2$IrO$_3$ estimated from a fit (not shown) of the $R_s(T)$ data to an activated Arrhenius behaviour.  

\begin{figure}[t]   
\includegraphics[width= 3 in]{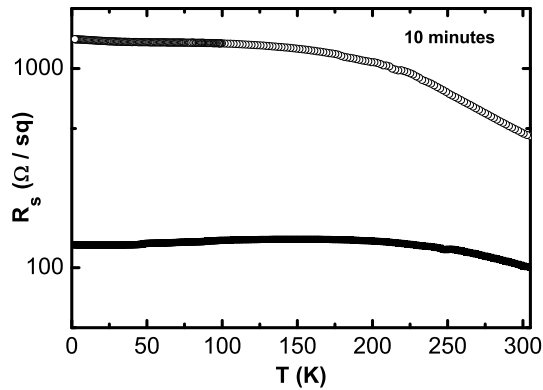}    
\caption{(Color online) A semi-log plot of the sheet resistance $R_s$ versus temperature $T$ for two samples of Na$_2$IrO$_3$ after $10$~minutes of plasma etching.
\label{Fig-rho-10}}
\end{figure}

Figure~\ref{Fig-rho-10} shows the $R_s(T)$ data for two single crystals etched for $10$~minutes.  While the $R_s$ values for the two samples are different by about an order of magnitude or less, the qualitative $T$ dependence is very similar with an increase on cooling from $T = 300$~K, a broad maximum reached around $150$--$200$~K, after which the $R_s$ is very weakly $T$ dependent.  For the sample with the lower resistance, the behaviour below the maximum is actually metallic.  

\begin{figure}[h]   
\includegraphics[width= 3 in]{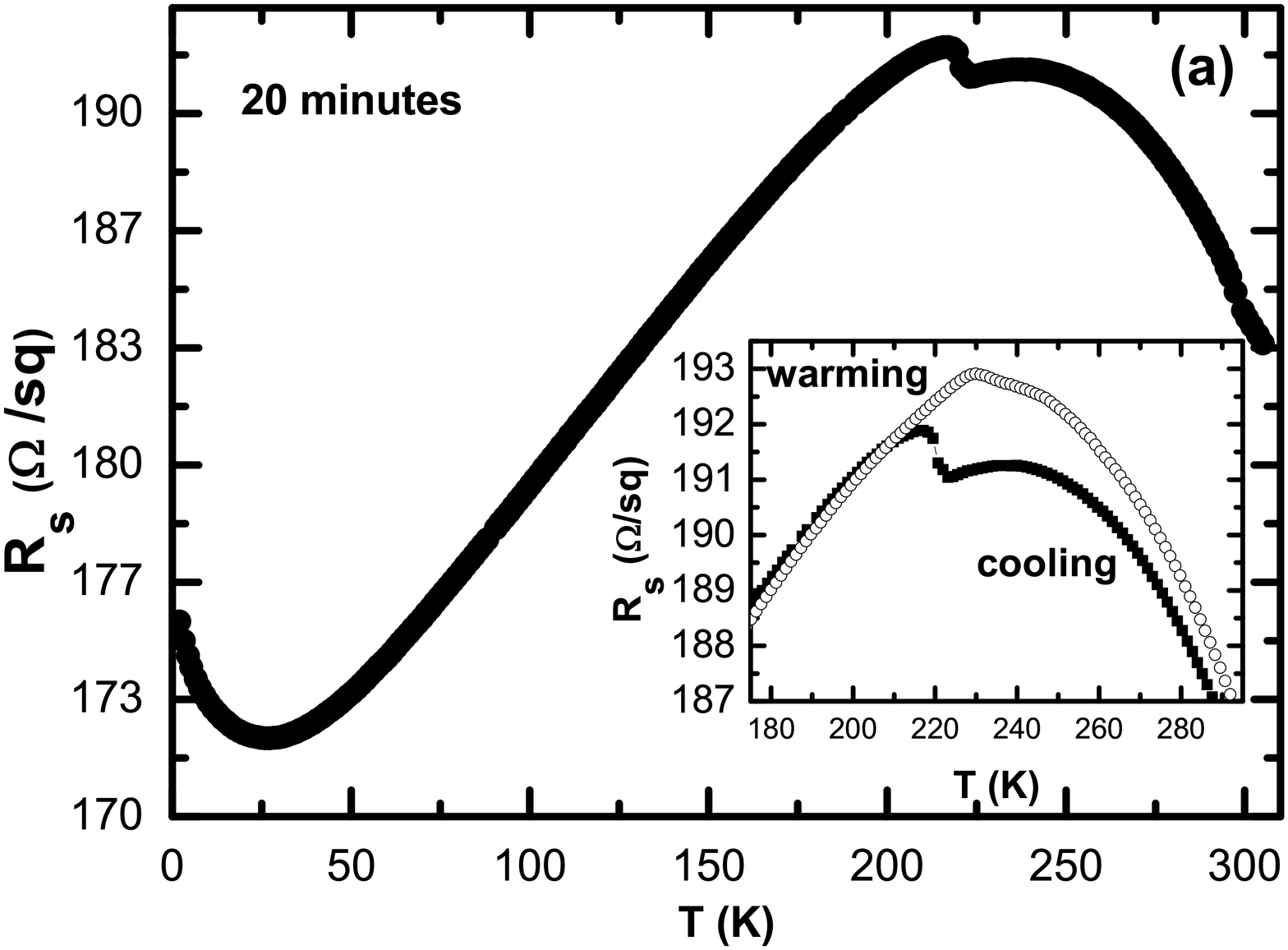} 
\includegraphics[width= 3 in]{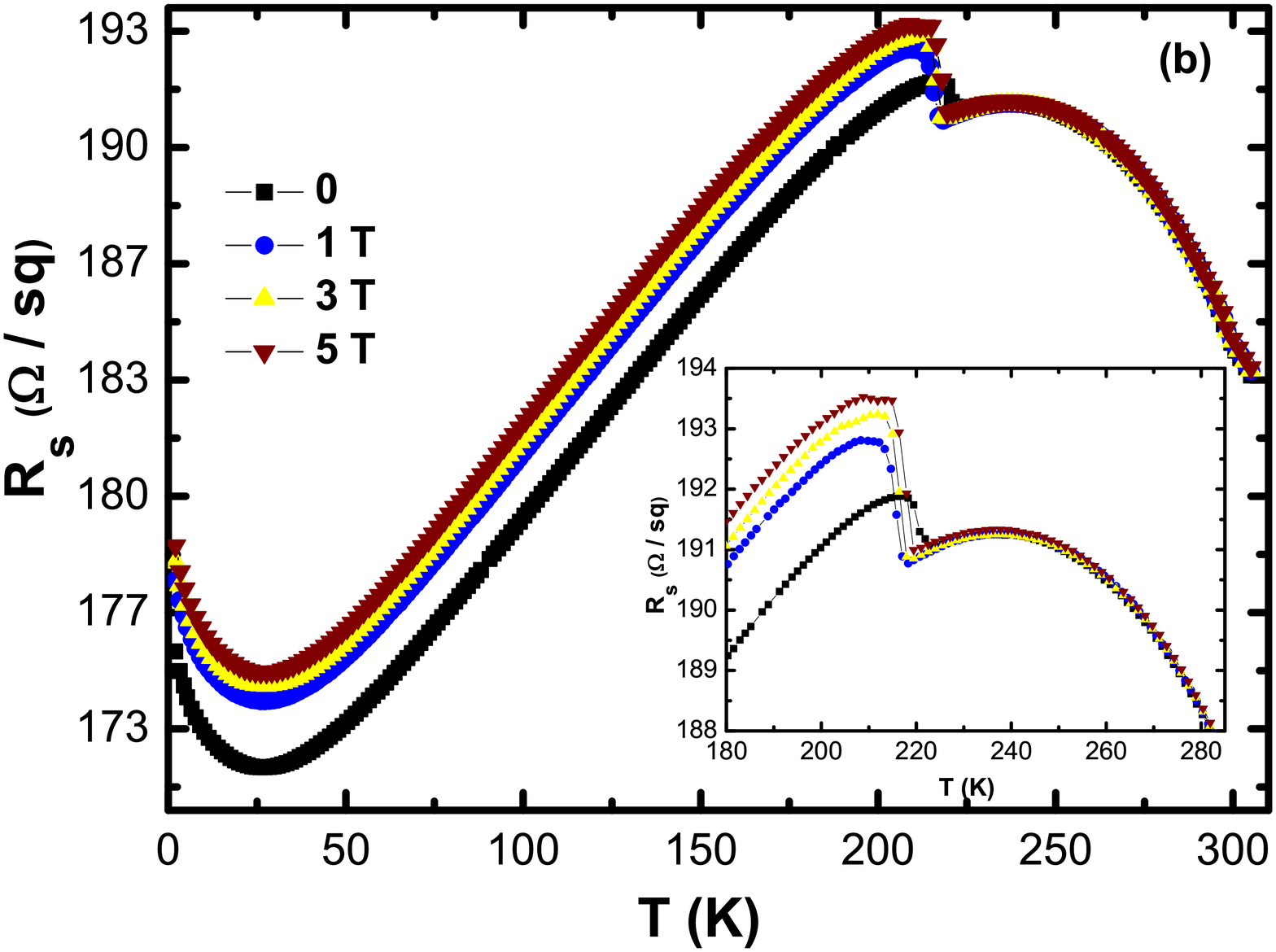}    
\caption{(Color online) Sheet resistance $R_s$ versus temperature $T$ for Na$_2$IrO$_3$ after $20$~minutes of Ar plasma etching.  (a) $R_s$ vs $T$ measured in zero magnetic field while cooling from $T= 305$~K\@.  Inset shows the cooling and warming data to highlight the thermal hysteresis indicating the first-order nature of the transition.  (b)  $R_s(T)$ measured while cooling from $T=305$~K in various applied magnetic fields $H$.  The inset shows the data close to the transition to highlight the field dependence at and below the transition.  
\label{Fig-rho-20}}
\end{figure}

Figure~\ref{Fig-rho-20}~(a) shows the zero magnetic field $R_s(T)$ data measured while cooling the sample exposed for $20$~minutes.  Resistance increases on cooling below $300$~K and reaches a maximum around $240$~K\@.  At $T_o \approx 220$~K, there is a transition involving an abrupt step-like increase in the resistivity.  Below $T_o$, metallic behavior is recovered down to $20$~K below which there is a slight increase in $R_s(T)$.  The step-like increase in $R_s(T)$ at $T_o$ is similar to the behavior observed for charge density wave transitions where the increase in $R_s$ results from a partial loss of density of states due to the opening up of a gap at the Fermi surface \cite{Gruener}.  The transition at $T_o$ is first-order as indicated by the hysteresis of about $10$~K between the cooling and warming curves shown in Fig.~\ref{Fig-rho-20}~(a) inset.  The transition is broadened out in the warming curve.  This behaviour was observed in repeated measurements and is intrinsic.  An additional feature which is not observed in conventional CDW systems is the magnetic field dependence.   Figure~\ref{Fig-rho-20}~(b) shows the $R_s(T)$ data measured while cooling in various magnetic fields $H$.  Above $T_o$ there is no $H$ dependence and all data fall on top of each other.  However, at and below $T_o$ there is an increase in the magnitude of $R_s(T)$ on the application of field although the qualitative behaviour remains the same.  The magnitude of the step-like increase in $R_s(T_o)$ also increases with $H$ as seen in Fig.~\ref{Fig-rho-20}~(b) inset.

\begin{figure}[t]   
\includegraphics[width= 3 in]{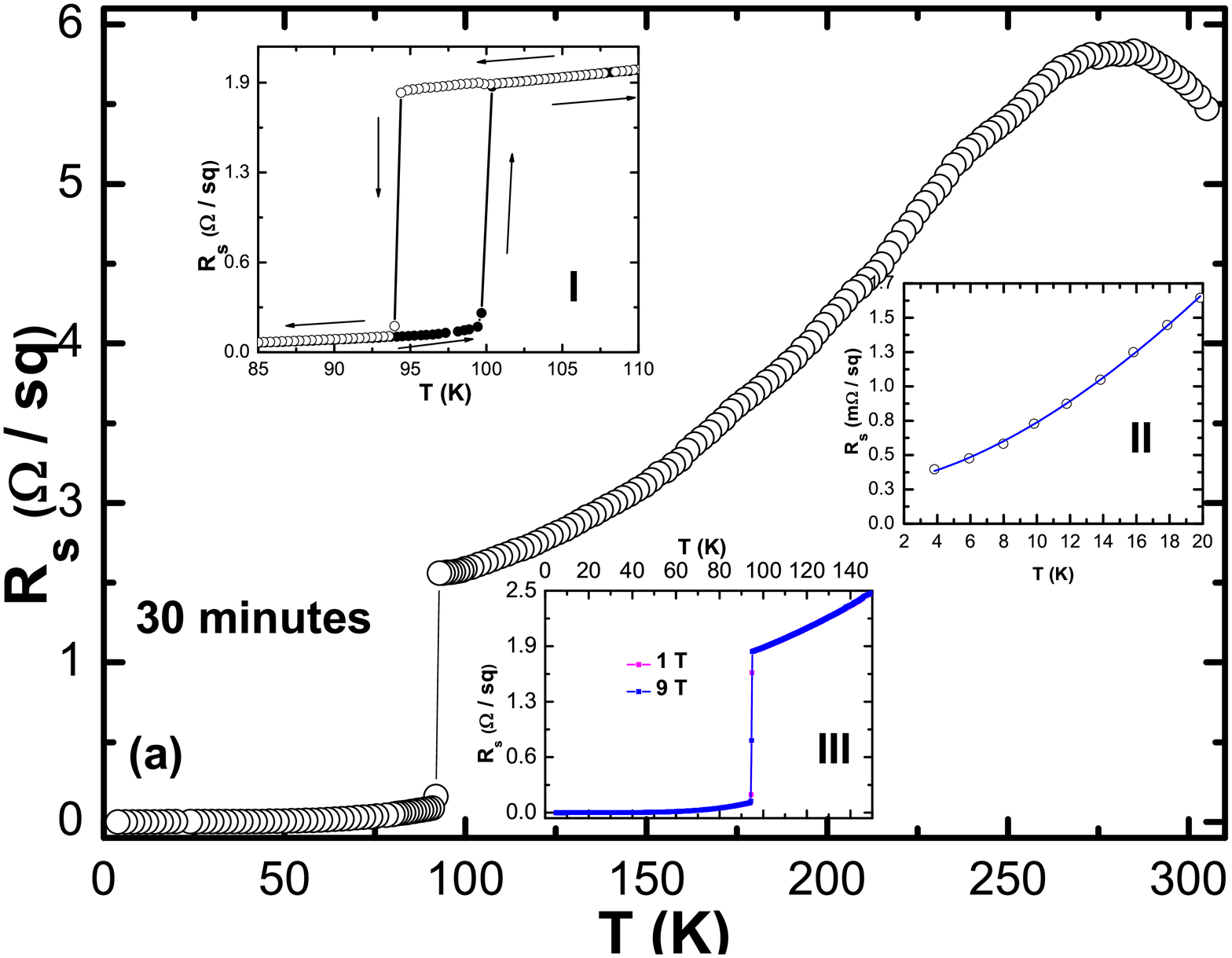}
\includegraphics[width= 3 in]{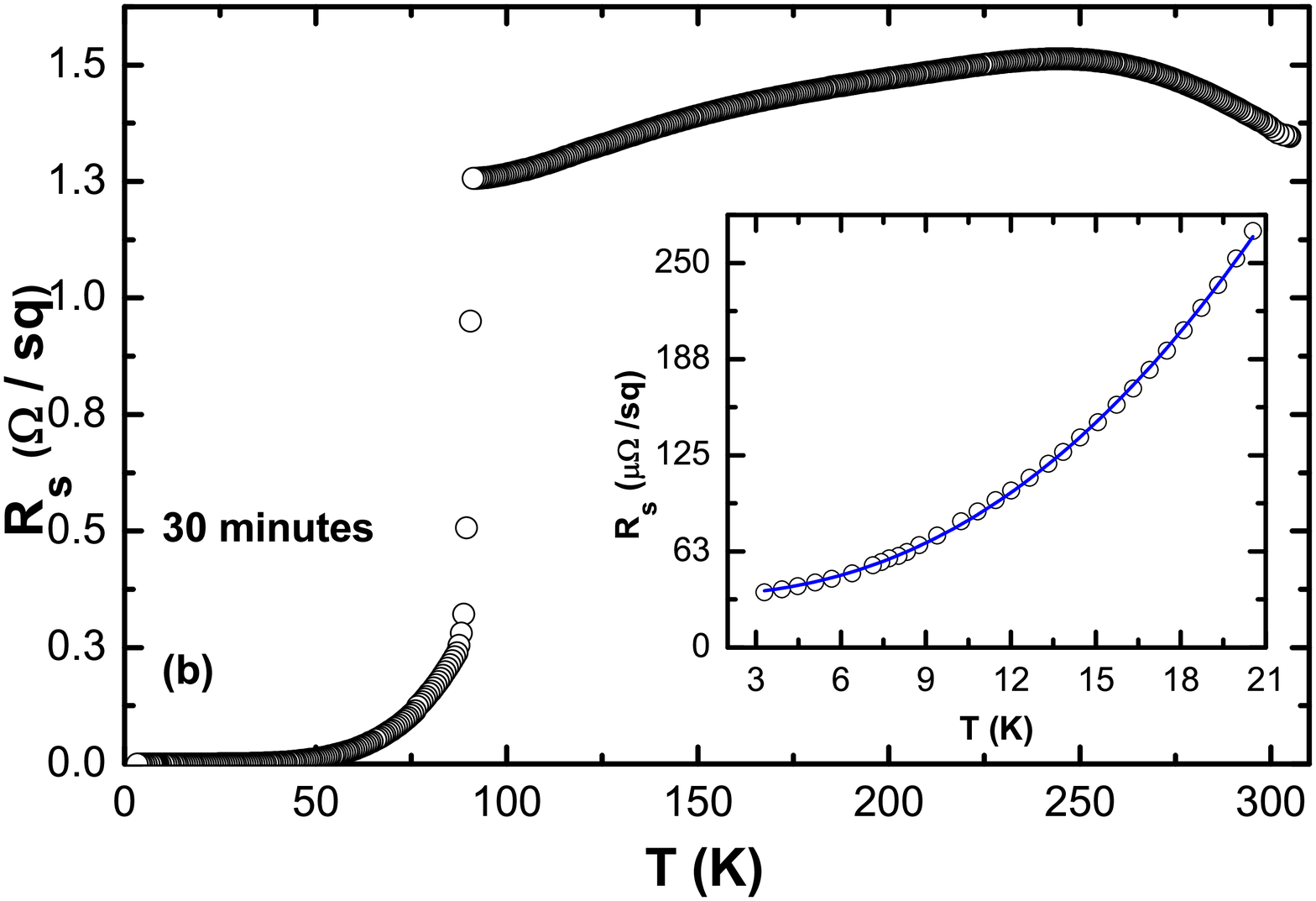}      
\caption{(Color online) Sheet resistance $R_s$ versus temperature $T$ for two Na$_2$IrO$_3$ Xtals after $30$~minutes of Ar plasma etching.  (a) $R_s$ vs $T$ of one Xtal measured in zero magnetic field while cooling from $T= 305$~K\@.  Inset~I shows the cooling and warming data to highlight the thermal hysteresis indicating the first-order nature of the transition.  Inset~II shows the low temperature data below $T = 20$~K\@.  The curve through the data is a fit to a $T^2$ dependence suggesting Fermi liquid behaviour.  Inset~III shows the $R_s(T)$ data measured in two magnetic fields to highlight the absence of any $H$ dependence. (b)  $R_s$ vs $T$ of the second Xtal measured in zero magnetic field while cooling from $T= 305$~K\@.  Inset shows the low temperature data below $T = 20$~K\@.  The curve through the data is a fit to a $T^2$ dependence suggesting Fermi liquid behaviour.
\label{Fig-rho-30}}
\end{figure}

Figures~\ref{Fig-rho-30}~(a) and ~(b) show the $R_s(T)$ data, measured while cooling, for two crystals exposed for $30$~minutes.  While the details are different, the qualitative behaviour for these two samples is the same.  On cooling one finds an increase in $R_s(T)$, a maximum around $250$--$275$~K, metallic behaviour down to $T_o \approx 95$~K where an abrupt step-like decrease in $R_s(T)$ by more than an order of magnitude occurs signalling a density wave-like transition.  This metal-to-metal transition is first order in nature as seen from the thermal hysteresis of about $10$~K between warming and cooling measurements shown in Fig.~\ref{Fig-rho-30}~(a)~inset~I\@.  The lowest temperature $R_s(T)$ data follow an approximately $T^2$ behaviour expected for a Fermi liquid metal.  This is shown in Fig.~\ref{Fig-rho-30}~(a)~inset~II and Fig.~\ref{Fig-rho-30}~(b)~inset, where the solid curve through the data are fits to a $T^2$ dependence.  Unlike for the $20$~minute sample, there is no magnetic field dependence of the resistivity for the $30$~minute sample as seen in Fig.~\ref{Fig-rho-30}~(a)~inset~III which shows the $R_s(T)$ data between $2$~K and $150$~K in a magnetic field of $1$~T and $9$~T\@.

\noindent
\emph{Summary and Discussion:} We have shown that the surface of Na$_2$IrO$_3$ crystals are highly tunable using plasma etching.  We were able to change the surface conductivity by several orders of magnitude by varying the plasma exposure time.  Specifically we studied samples irradiated for times $t = 0, 5, 10, 20, 30$~minutes.  GISAXS measurements showed that the surface structure does not change after the etching and EDS chemical analysis revealed that Na was being progressively removed on increasing the etching time.  The samples etched for $10$~minutes showed unusual transport behavior with an increase in resistivity $R_s$ on cooling from $300$~K down to $150$--$200$~K below which $R_s(T)$ became almost $T$-independent (see Fig.~\ref{Fig-rho-10}).  This behavior is consistent with that expected for topological insulators where at low temperatures surface conductivity starts contributing after the bulk becomes sufficiently insulating.  It must be noted that topological insulating state has been predicted for Na$_2$IrO$_3$ under certain conditions \cite{Shitade2009, Kim2013}.  Whether this behaviour in doped Na$_2$IrO$_3$ has any topological properties will have to be explored in more detailed surface sensitive studies in future.  

The samples etched for $20$ and $30$ minutes show bad-metal behavior at high temperature with large values ($\sim \Omega$) and with an increasing $R_s(T)$ with decreasing $T$.  The $R_s(T)$ passes over a maximum around $250$~K and turns metallic for lower temperatures.  For the $20$($30$)~minute samples, we observe abrupt step-like increase(decrease) in $R_s(T)$ at $T_o \approx 220 (95)$~K\@.  These are first-order transitions as revealed by a $10$~K thermal hysteresis.  These signatures in electrical transport are reminiscent of spin or charge density wave transitions \cite{Gruener, Yang1991, Coleman1985, Brooks2006, Singh2005, Ramakrishnan2007, Ong1977, Thompson1972, Friend1977, Galli2000, Tiedje1975, Fawcett1988}.  Usually, at a charge density wave transition, a periodic lattice distortion is accompanied by a (partial) gapping of the Fermi surface.  This leads to an increased resistance below the CDW transition \cite{Gruener, Yang1991, Singh2005, Ong1977}.  This is consistent with the feature observed at $\approx 220$~K for the $20$~minute etched sample.  However, a drop in resistivity, as seen at $\approx 95$~K for the $30$~minute etched sample, is hard to explain as arising from a CDW transition with a gapping mechanism.  Nevertheless, prominent examples of this behaviour are known to exist. For example, the resistivity for 2H- TaS$_2$ at the CDW transition at $75$~K \cite{Thompson1972}, for 4H-TaS$_2$ at the CDW transition at $22$~K \cite{Friend1977}, and for Er$_5$Ir$_4$Si$_{10}$ at the CDW transition at $55$~K \cite{Galli2000}, drop abruptly. The mechanism leading to the drop in resistivity at the CDW transition not understood fully although it was speculated in the case of Er$_5$Ir$_4$Si$_{10}$ that the lattice modulation led to a band structure which led to enhanced conductivity \cite{Galli2000}. The CDW transitions accompanied by periodic lattice distortions in all these materials have been verified and demonstrated through electron diffraction measurements.  

The magnetic field dependence of the two proposed density wave-like transitions are also different.  While the resistance at and below the transition at $220$~K for the $20$~minute sample is field dependent, there is no effect of the field on the resistance of the $30$~minute etched sample across the transition at $95$~K\@.  There are again examples of both kind of behaviours observed for well established spin and charge density wave materials and the details of the electronic structure, electron-phonon coupling, and DW instability mechanism govern whether a field will have any effect on the transport properties or not.  For example, the resistance at and below the lower CDW transition in NbSe$_3$ is strongly enhanced by a magnetic field and even the transition temperature is increased by a small amount \cite{Coleman1985}.  This has been suggested to happen because the magnetic field improves the imperfect nesting which already existed at zero field.  The improved nesting in an applied field leads to a larger portion of the Fermi surface becoming gapped and hence leads to a larger resistance below the CDW transition \cite{Coleman1985}.  On the other hand, in the 1-dimensional organic conductor TTF-TCNQ, the resistance at and below the $58$~K CDW transition and indeed the transition temperature itself is suppressed on the application of a magnetic field \cite{Tiedje1975}.  This can be understood in terms of the competition of the Zeeman splitting of the nested bands in a field and the CDW gap in analogy to the breaking of Cooper pairs in BCS superconductivity \cite{Dieterich1973}.  Similar suppression of CDW with magnetic field has been reported for another 1-D organic conductor Per$_2$[Au(mnt)$_2$] \cite{Graf2004}. Additionally, there could be magnetic field effects at a CDW transition if there is a coupled SDW state occurring simultaneously.  This is seen at the SDW transition in metallic Chromium \cite{Fawcett1988}.

Thus, the transport anomalies observed for doped Na$_2$IrO$_3$ are consistent with spin or charge density wave transitions.  Remarkably, such spin and charge density waves, spin/charge bond-order, and structural instabilities have been predicted for the doped Kitaev-Heisenberg model relevant for Na$_2$IrO$_3$ \cite{Scherer2014}.  Further experimental work and in particular surface sensitive probes like STM would be useful to reveal the nature and origin of the remarkable features seen in doped Na$_2$IrO$_3$. 
        
\noindent
\emph{Acknowledgments.--} We thank the small angle x-ray scattering facility at IISER Mohali. We thank Dr. A. Venkatesan for use of the plasma etching facility and we acknowledge the SEM facility at IISER Mohali for chemical analysis.  KM acknowledges UGC-CSIR India for a fellowship.  YS acknowledges DST, India for support through Ramanujan Grant \#SR/S2/RJN-76/2010 and through DST grant \#SB/S2/CMP-001/2013.

\end{document}